\documentclass[twocolumn,english,amsmath,amssymb,aps,prl,showpacs]{revtex4}
\usepackage[T1]{fontenc}
\usepackage[latin9]{inputenc}
\usepackage{color}
\usepackage{babel}
\usepackage{mathrsfs}
\usepackage{amsmath}
\usepackage{amssymb}
\usepackage{graphicx}
\usepackage{esint}
\usepackage[unicode=true,pdfusetitle,
 bookmarks=false,
 breaklinks=false,pdfborder={0 0 1},backref=false,colorlinks=true]
 {hyperref}
\hypersetup{
 colorlinks,linkcolor=blue,citecolor=blue,urlcolor=blue}

\makeatletter

\newcommand*\LyXThinSpace{\,\hspace{0pt}}

\@ifundefined{textcolor}{}
{%
 \definecolor{BLACK}{gray}{0}
 \definecolor{WHITE}{gray}{1}
 \definecolor{RED}{rgb}{1,0,0}
 \definecolor{GREEN}{rgb}{0,1,0}
 \definecolor{BLUE}{rgb}{0,0,1}
 \definecolor{CYAN}{cmyk}{1,0,0,0}
 \definecolor{MAGENTA}{cmyk}{0,1,0,0}
 \definecolor{YELLOW}{cmyk}{0,0,1,0}
}

\makeatother

\begin{document}

\title{Crossover to the Anomalous Quantum Regime in the Extrinsic Spin Hall
Effect of Graphene}

\author{Mirco Milletar\`{i}}
\email{milletari@gmail.com}

\affiliation{Centre for Advanced 2D Materials and Department of Physics, National
University of Singapore, Singapore, 117551}

\author{Aires Ferreira}
\email{aires.ferreira@york.ac.uk}

\affiliation{Department of Physics, University of York, York YO10 5DD, United
Kingdom}
\begin{abstract}
Recent reports of spin\textendash orbit coupling enhancement in chemically
modified graphene have opened doors to studies of the spin Hall effect
with massless chiral fermions. Here, we theoretically investigate
the interaction and impurity density dependence of the extrinsic spin
Hall effect in spin\textendash orbit coupled graphene. We present
a nonperturbative quantum diagrammatic calculation of the spin Hall
response function in the strong-coupling regime that incorporates
skew scattering and \emph{anomalous} impurity density-independent
contributions on equal footing. The spin Hall conductivity dependence
on Fermi energy and electron\textendash impurity interaction strength
reveals the existence of experimentally accessible regions where anomalous
quantum processes dominate. Our findings suggest that spin\textendash orbit-coupled
graphene is an ideal model system for probing the competition between
semiclassical and \emph{bona fide} quantum scattering mechanisms underlying
the spin Hall effect. 
\end{abstract}


\maketitle
Spintronics aims to explore charge, spin and orbital degrees of freedom
of electrons to realize novel approaches to advanced storage and logic
computing \cite{Wolf_01}. Graphene\textemdash a one-atom thick layer
of carbon atoms with unique electronic properties \cite{Graph_advent}\textemdash holds
promising applications in spintronics \cite{Han_14}. The weak spin\textendash orbit
coupling \cite{Hernando06,Fabian09} and high mobilities of $sp_{2}$-hybridized
carbon result in large spin diffusion lengths (e.g., 1\textendash 20\,$\mu$m
in exfoliated samples \cite{Tombros_07,Zomer_12}), making graphenic
systems attractive as spin channels of high performance \cite{Tombros_07,Zomer_12,Han_10}.

Recent progress in engineering of enhanced spin-orbit coupling (SOC)
in graphene through addition of impurities \cite{SOC_G_Balakrishnan_13,SOC_G_Balakrishnan_14}
and via coupling to suitable substrates \cite{SOC_G_Marchenko_13,SOC_G_Avsar_14,SOC_G_Wang_Morpurgo_15,SOC_G_Mendes_15}
opens up intriguing possibilities. The presence of spin\textendash orbit
interactions is predicted to profoundly alter the standard pictures
of spin relaxation \cite{Federov_13,Bundesmann_15} and weak localization
\cite{McCann12}. Furthermore, a sizable SOC enables spin-dependent
transport phenomena absent in pristine samples \cite{Ferreira14,Pachoud,Barnas2011,SOC_G_Th_Asmar_13_15,SOC_G_Th_Dyrdal},
most noticeably the spin Hall effect (SHE), whereby charge currents
driven by electric fields are converted to transverse spin currents
\cite{SHE_Dyakanov,SHE_Hirsch,SHE_Zhang}. This phenomenon was first
observed by optical means in semiconductors in 2004 \cite{SHE_Kato,SHE_Wunderlich},
and its reciprocal\textemdash the inverse SHE\textemdash just shortly
after demonstrated by direct electrical measurements in metals \cite{SHE_Saitoh_06,SHE_Valenzuela_06}.
According to theory, a modest SOC in the range of 10 meV in graphene
enables robust and gate-tunable SHE \cite{Ferreira14}. Recent reports
on SHE exploring Hanle precession in adatom-decorated graphene \cite{SOC_G_Balakrishnan_13,SOC_G_Balakrishnan_14}
and graphene\textendash WS$_{2}$ heterostructures \cite{SOC_G_Avsar_14,SOC_G_Wang_Morpurgo_15},
and spin pumping in graphene/YIG devices \cite{SOC_G_Mendes_15},
confirm theoretical predictions, and pave the way for \emph{all electric}
spintronics in graphene.

Generally, two types of SHE can occur in a spin\textendash orbit-coupled
graphene system. When charge carriers experience a \emph{global} SOC\textemdash endowed
by proximity effect\textemdash a SHE is induced by the Berry curvature
of Bloch bands (the so-called ``intrinsic mechanism''), with scattering-dependent
corrections due to disorder \cite{Bruno_01}. Conversely, if the SOC
enhancement is confined to random ``hot spots''\textemdash e.g.,
as mediated by impurities\textemdash two basic mechanisms can compete
to establish a SHE, \emph{viz.}, the left/right asymmetric (skew)
scattering for spin-up and spin-down electrons \cite{Ferreira14,Pachoud},
and the quantum side-jump (QSJ) effect. The latter can be viewed as
a coordinate shift of wavepackets upon scattering in the presence
of SOC. The side jump is transverse to the external electric field
and has opposite signs for spin-up/down electrons, which results in
a net contribution to the spin Hall conductivity \cite{Bruno_01,Levy_88,Synitsin_review,Nagaosa_review,Levy_13}.

Owing to the sharpness of resonant scattering characteristic of massless
fermions in 2D \cite{RS_Stauber,RS_Robinson,RS_Wehling,RS_Ferreira},
the extrinsic SHE induced by skew scattering from SOC-active impurities
in graphene is predicted to be extremely robust, capable of yielding
giant spin Hall angles of the order of $0.1$ \cite{Ferreira14,Pachoud,Cazalilla16}.
For a very low concentration of impurities, quantum contributions
to the spin Hall (SH) conductivity are negligible, and the semiclassical
skew scattering fully determines the steady state of SHE \cite{Ferreira14}.
However, much less is known about the role of quantum processes in
the dilute regime of much interest in extrinsic graphene ($\approx0.01-0.1\%$
atomic ratio \cite{SOC_G_Balakrishnan_13,SOC_G_Balakrishnan_14,Adatoms}),
especially in the strong scattering limit, where quantum contributions
to the SH response functions are hard to assess \cite{Rammer}.

In this paper, we present a microscopic theory of the extrinsic SHE
in graphene based on a\emph{ }nonperturbative quantum diagrammatic
calculation able to capture the strong scattering regime self-consistently.
We find that skew scattering, QSJ, and multiple impurity scattering
processes need to be considered \emph{on equal footing} for an accurate
description of the extrinsic SHE. Quite remarkably, a crossover towards
an ``anomalous phase''\textemdash where quantum processes overcome
skew scattering\textemdash is shown to occur in experimentally accessible
parameter regions. Our self-consistent approach goes beyond previous
theories \cite{Ferreira14,Bruno_01,Levy_13,Levy_88,SHE_Zhang}, providing
a unified description of skew scattering and side jump mechanisms.

\begin{figure}[t!]
\centering{}\includegraphics[width=0.8\columnwidth]{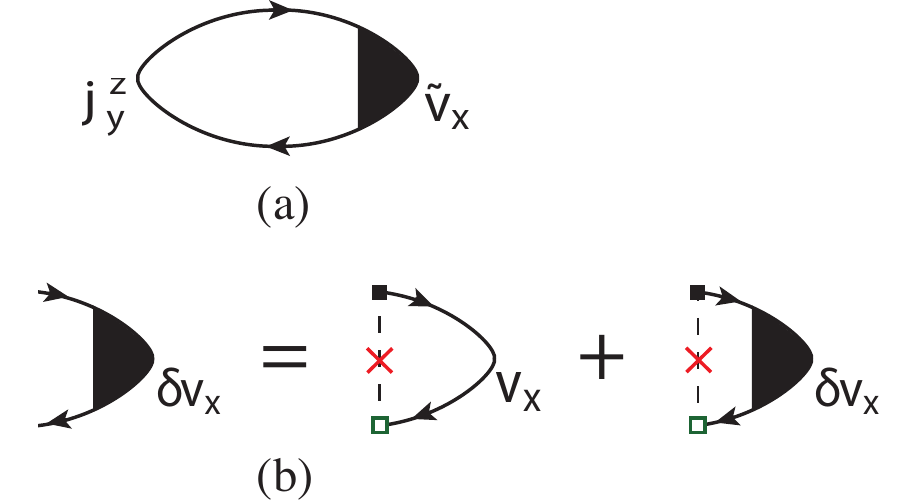} \caption{Kubo\textendash Streda diagrams. (a) Response bubble for the SH conductivity
with dressed charge vertex $\tilde{v}_{x}=v_{x}+\delta v_{x}$. (b)
Bethe-Salpeter equation for the vertex correction $\delta v_{x}$.}
\label{fig:vertexc} 
\end{figure}

\emph{Model system.\textemdash }The low-energy physics of spin\textendash orbit-coupled
graphene is described by a Dirac Hamiltonian in two spatial dimensions
with a random impurity potential. For simplicity, the typical size
of SOC-active impurities is assumed much larger than the lattice spacing,
hence suppressing intervalley scattering \cite{Ferreira14,Pachoud}.
We work with the SO(5) representation of the spin algebra \cite{Murakami,Zee}
in terms of $4\times4=1+5+10$ matrices, i.e., one identity, $\gamma^{0}$,
five $\gamma^{a}$ matrices, taken as $\gamma^{1}=\sigma_{1}\otimes s_{0}$,
$\gamma^{2}=\sigma_{2}\otimes s_{0}$, $\gamma^{3}=\sigma_{3}\otimes s_{3}$,
$\gamma^{4}=\sigma_{3}\otimes s_{2}$, and $\gamma^{5}=\sigma_{3}\otimes s_{1}$,
and ten adjoint matrices $\gamma^{ab}=i/2\,[\gamma^{a},\gamma^{b}]$.
Here $\boldsymbol{\sigma}$ and $\boldsymbol{s}$ are Pauli matrices
defined in the sublattice and spin space, respectively. The Hamiltonian
density reads\textcolor{blue}{{} }\textcolor{black}{{} 
\begin{equation}
\mathscr{H}=\psi^{\dagger}(\mathbf{x})\left\{ -i\,v\,\gamma^{j}\partial_{j}-\gamma_{0}\,\epsilon+V(\mathbf{x})\right\} \psi(\mathbf{x}),\label{eq:Ldirac}
\end{equation}
}where $v$ is the Fermi velocity of charge carriers, $\epsilon$
is the Fermi energy, and $V(\mathbf{x})$ denotes the disorder potential.
Hereafter, we set $\hslash\equiv1\equiv e$, unless stated otherwise.
The impurities are modeled as short-range potentials, $V(\mathbf{x})=\sum_{i=1}^{N}M\,R^{2}\delta(\mathbf{x}-\mathbf{x}_{i})$,
where $M$ is a $4\times4$ matrix encoding the spin and sublattice
structure of the impurity, and $R$ is a length scale mimicking a
potential range \cite{RS_Ferreira}. We posit our analysis on impurities
leading to a SOC of the \emph{``}intrinsic type'' \cite{Hernando06,Fabian09}
and allow for an extra (scalar) electrostatic term in the impurity
matrix: 
\begin{equation}
M=\alpha_{0}\,\gamma_{0}+\alpha_{3}\,\gamma_{3}\,,\label{eq:M_matrix_intrinsic_SOC}
\end{equation}
with $\alpha_{0}$ ($\alpha_{3}$) denoting the magnitude of the scalar
(SOC) component of the disordered potential. Note that $\gamma_{3}$
conserves the out-of-plane spin component, in addition to being an
invariant of the $C_{6v}$ point group, and thus is the simplest form
of SOC in graphene; physical realizations include physisorbed atoms
in the hollow position, and top-position adatoms randomly distributed
over sublattices \cite{Pachoud,Fabian_H_F}. 
\begin{figure}[t!]
\centering{}\includegraphics[width=0.95\columnwidth]{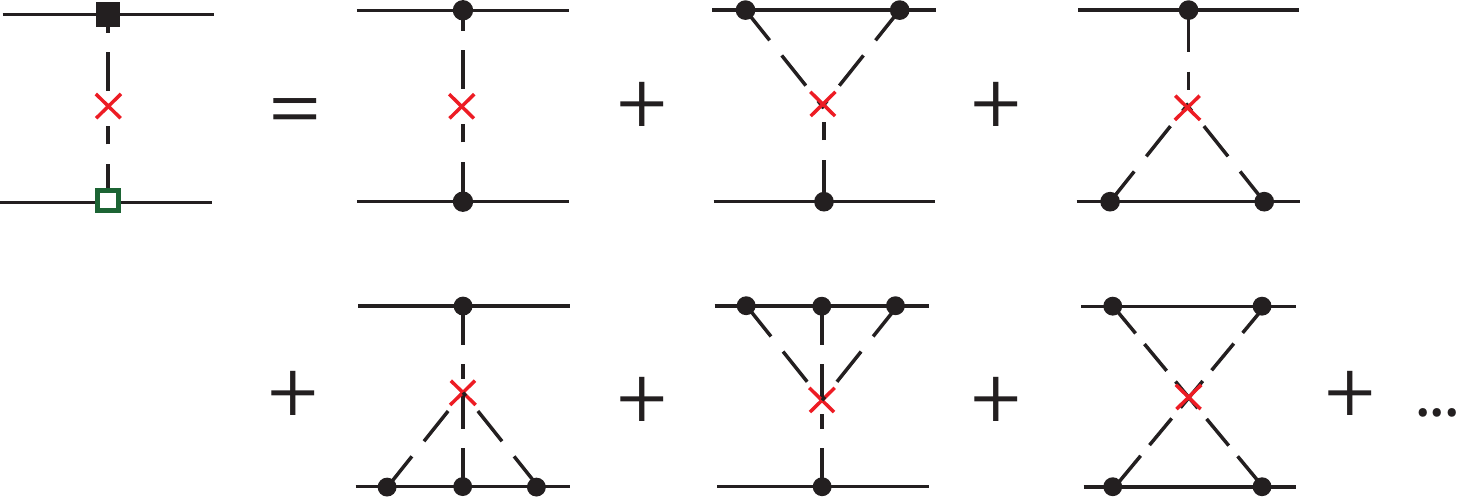} \caption{$T$~Matrix ladder. Skeleton expansion of the ladder diagram in terms
of an infinite series of two particle, noncrossing diagrams. On the
left side, a full (open) square interaction vertex denotes a $T$
($T^{*}$)~matrix insertion, while on the right the $T$~matrix
is expanded in its bare components ($M$ insertions). The red $\times$
represents an impurity density insertion. }
\label{fig:T-ladder} 
\end{figure}

\emph{Methodology}.\textemdash \textcolor{black}{Being interested
in the effect of asymmetric and strong scattering,} the standard \emph{Gaussian
white noise} approximation is not applicable. Instead, we employ the
$T$-matrix approach valid for a low density of impurities with otherwise
arbitrarily strong scattering potential. The $T$~matrix is the result
of an infinite order resummation of potential scattering diagrams
containing only one impurity density insertion $n=N/\Omega$ (here
$\Omega$ is the sample area) in the non-crossing approximation~\cite{Rammer}.
The self energy reads $\Sigma(\epsilon)=n\,\langle T(\epsilon)\rangle_{\textrm{dis}}$,
where $\langle...\rangle_{\textrm{dis}}$ denotes configurational
average. We find, after ressumation, $\langle T(\epsilon)\rangle_{\textrm{dis}}=\frac{1}{2}\left(T_{+}+T_{-}\right)\gamma_{0}+\frac{1}{2}\left(T_{+}-T_{-}\right)\gamma_{3}\equiv T$,
with 
\begin{equation}
T_{\pm}=\frac{R^{2}\,(\alpha_{0}\pm\alpha_{3})}{1-R^{2}\,(\alpha_{0}\pm\alpha_{3})\,g_{0}(\epsilon)}\equiv\epsilon_{\pm}\mp i\,\eta_{\pm}.\label{eq:selfE}
\end{equation}
In the above, $g_{0}(\epsilon)=-|\epsilon|/2\pi v^{2}\textrm{ln}\left(\Lambda/|\epsilon|\right)\mp i\,|\epsilon|/4v^{2}$
is the momentum integrated bare propagator in retarded (advanced)
sectors, and $\Lambda$ is a high energy cutoff \cite{RS_Ferreira}.
To simplify notation, hereafter $\epsilon\ge0$ is assumed. It is
convenient to decompose the self energy in real and imaginary part
as: $\Re\,\Sigma=n(\delta\epsilon\,\gamma_{0}+m\,\gamma_{3})$ and
$-\Im\,\Sigma=n(\eta\,\gamma_{0}+\bar{\eta}\,\gamma_{3})$, where
$\delta\epsilon=(\epsilon_{+}+\epsilon_{-})/2$, $m=(\epsilon_{+}-\epsilon_{-})/2$,
$\eta=(\eta_{+}+\eta_{-})/2$ and $\bar{\eta}=(\eta_{+}-\eta_{-})/2$.
Here, $n\,\delta\epsilon$ is a chemical potential shift that can
be reabsorbed in $\epsilon$, while $n\,m$ is a (small) disorder-induced
SOC gap. This result shows that $\hat{\Sigma}$ endows quasiparticles
with two different lifetimes; we have defined $n\,\eta$ and $n\,\bar{\eta}$
as the respective energy and spin gap broadenings. The disorder averaged
propagator reads 
\begin{equation}
\mathcal{G}_{\mathbf{k}}^{R/A}(\epsilon)=\frac{(\epsilon\pm i\,n\,\eta)\gamma_{0}+n\,(m\mp i\,\bar{\eta})\gamma_{3}+v\,\gamma^{j}k_{j}}{(\epsilon\pm i\,n\,\eta)^{2}-n^{2}(m\mp i\,\bar{\eta})^{2}-v^{2}\,k^{2}}.\label{eq:avProp}
\end{equation}
It is interesting to note that the above propagator has a structure
similar to that found in minimal models of the anomalous Hall effect
(AHE) based on the massive Dirac equation in $d=2+1$ \cite{Synitsin_link,Ado}
(note, however, the physically distinct origins of the respective
$\gamma_{3}$ ``mass'' terms). Next, we evaluate the SH conductivity
using the Kubo\textendash Streda formula, represented diagrammatically
in Fig.~\eqref{fig:vertexc}. In our model, the spin and charge vertex
are given, respectively, by $j_{y}^{z}=v/2\,\gamma_{13}$ and $v_{x}=v\,\gamma_{1}$.

\emph{Bubble} \emph{approximation; unitary vs Gaussian limits}.\textemdash It
is instructive to first consider the limiting cases of infinitely
strong (unitary) and weak (Gaussian) scatterers. Neglecting the vertex
corrections for the moment, we obtain to leading order in the impurity
density, and including a valley degeneracy factor of two: 
\begin{equation}
\sigma_{\textrm{SH}}^{0}=2\int\frac{d^{2}\mathbf{k}}{(2\pi)^{2}}\:\textrm{Tr}\left[\,j_{y}^{z}\,\mathcal{G}_{\mathbf{k}}^{R}(\epsilon)\,v_{x}\,\mathcal{G}_{\mathbf{k}}^{A}(\epsilon)\,\right]\simeq\frac{\bar{\eta}}{\eta}.\label{eq:bubble}
\end{equation}
The bubble SH conductivity is a ratio of two broadening scales and
hence is independent on the impurity density; the underlying SH mechanism
is the QSJ \cite{Synitsin_review}. In the \emph{unitary} limit, $|\Re\,g_{0}\,R^{2}(\alpha_{0}\pm\alpha_{3})|\gg1$,
$\eta_{\pm}\approx\pi^{2}v^{2}/\epsilon\textrm{ln}(\Lambda/\epsilon)$,
and hence the SH conductivity is identically zero. On the other hand,
in the \emph{Gaussian} limit, $|\Re\,g_{0}\,R^{2}(\alpha_{0}\pm\alpha_{3})|\ll1$,
$\eta_{\pm}\simeq R^{4}(\alpha_{0}\pm\alpha_{3})^{2}\epsilon/(4v^{2})$,
and one obtains a non-zero result, $\sigma_{\textrm{SH}}^{0}=\alpha_{0}\,\alpha_{3}/(\alpha_{0}^{2}+\alpha_{3}^{2})$.
The Gaussian approximation then gives an energy independent contribution,
while dependence on the Fermi energy only appears at order $n$ and
it is therefore sub-leading in the dilute regime. However, a careful
analysis shows that this result is an artifact of the Gaussian approximation.
In order to obtain the correct dependence on the Fermi energy, a calculation
based on the full $T$~matrix approach is required.

\begin{figure}[!t]
\begin{centering}
\includegraphics[width=0.9\linewidth]{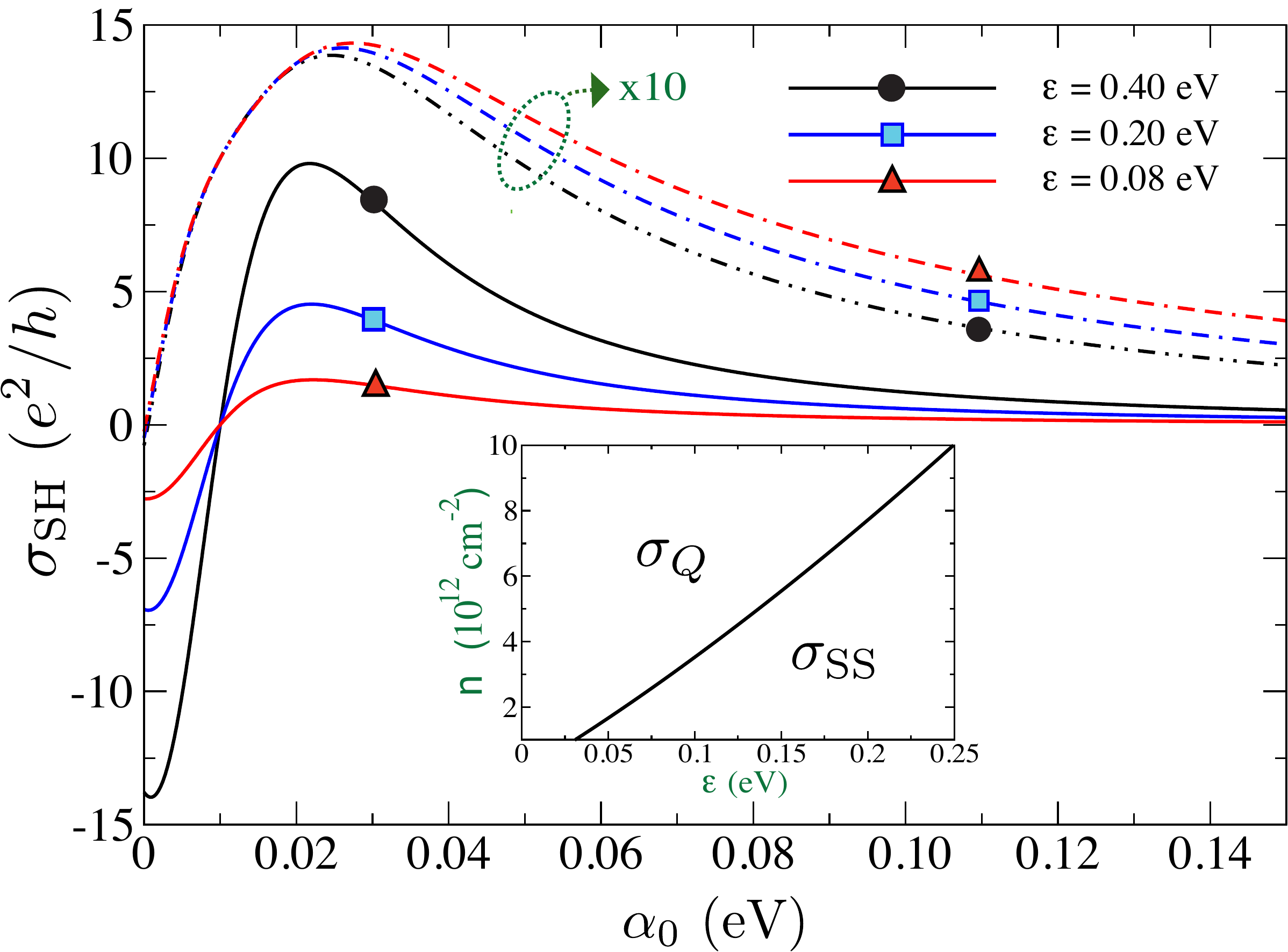} 
\par\end{centering}
\centering{}\caption{SH conductivity. The semiclassical SS and anomalous contributions
to $\sigma_{\textrm{SH}}$ are shown for different values of the Fermi
energy in solid and dotted lines, respectively. $\sigma_{\textrm{SS}}$
($\sigma_{Q}$) increases (decreases) with $\epsilon$, and both conductivities
decrease at increasing scalar potential magnitude, in agreement with
the unitary limit result. Note that $\sigma_{Q}$ has been scaled
by a factor of 10. We have used $\alpha_{3}=0.01$~eV, $R=4$~nm,
and $n=4\times10^{12}$~cm$^{-2}$, typical parameters for physisorbed
metal nanoparticles \cite{SOC_G_Balakrishnan_14,Ferreira14}. The
inset shows the regions $(\epsilon,n)$ dominated by the semiclassical
and anomalous contributions ($\alpha_{0}=0.05$ eV, other parameters
as in main figure). }
\label{fig:sigmaSH} 
\end{figure}

\emph{Full calculation}.\textemdash The $T$~matrix enters the problem
in the propagators (via self energy) and in the response bubble itself
(4-point function). The former has already been evaluated below Eq.~\eqref{eq:selfE},
we now tackle the 4-point function. Figure~\eqref{fig:T-ladder}
shows the dressed ladder diagram and its skeleton expansion. In order
to describe the strong scattering regime, one needs to change the
Feynman's rules for disorder potential insertions from the standard
bare interaction (dot) to the $T$~matrix-dressed one (squares).
This procedure generates \emph{all diagrams} with one impurity density
insertion (one $\times$), thus providing an accurate nonperturbative
result. The treatment of 4-point electron-hole propagators at the
$T$-matrix level has been employed in Ref.~\cite{Hirsch} in the
context of resonant scattering in anisotropic superconductors. Although
previously neglected in studies of anomalous and SH effects, the additional
(4-point) diagrams are essential to describe the strong scattering
regime relevant for SHE in spin\textendash orbit-coupled graphene.
In the skeleton expansion of Fig.~\eqref{fig:T-ladder}, one recognizes
the first term as the bare ladder diagram, providing the first correction
to the empty bubble, Eq.~\eqref{eq:bubble}. The next two diagrams
in the figure (``$Y$ diagrams'') contain three $M$ impurity insertions,
and hence encode skew scattering (SS) at the lowest order \cite{Bruno_01,Synitsin_review,Synitsin_link,Milletari_2}.
The remaining diagrams build up the complete 4-point skeleton series
describing QSJ and SS processes at all orders in the impurity potential.

The charge vertex is schematically shown in Fig.~\eqref{fig:vertexc},
together with the conductivity diagram. We first evaluate the single-impurity
vertex correction $\bar{v}_{x}$. Using the $T$~matrix ladder diagram
shown in Fig.~\eqref{fig:T-ladder}, we find 
\begin{align}
\bar{v}_{x} & =n\,\int\frac{d^{2}\mathbf{k}}{(2\pi)^{2}}\:T\,\mathcal{G}_{\mathbf{k}}^{R}\,v_{x}\,\mathcal{G}_{\mathbf{k}}^{A}\,T^{*}=v\,(a\,\gamma_{1}+b\,\gamma_{13})\,,\label{eq:Tladder}\\
a & \simeq\epsilon\,\frac{\eta_{+}\,\eta_{-}+\epsilon_{+}\,\epsilon_{-}}{4v^{2}(\eta_{+}+\eta_{-})}-n\,f_{a}(\eta_{+},\eta_{-},\epsilon_{+},\epsilon_{-}),\nonumber \\
b & \simeq\epsilon\,\frac{\eta_{+}\,\epsilon_{-}-\eta_{-}\,\epsilon_{+}}{4v^{2}(\eta_{+}+\eta_{-})}+n\,f_{b}(\eta_{+},\eta_{-},\epsilon_{+},\epsilon_{-}),\nonumber 
\end{align}
where $f_{a}$ and $f_{b}$ are complicated functions of $\eta_{\pm},\epsilon_{\pm}$;
explicit expressions are given in the Supplemental Material (SM) \cite{SM}.
Note that contrary to the Gaussian case, also $b$ starts constant
in $n$. This term is responsible for the semiclassical SS, yielding
the standard skew relaxation-time contribution, $\sigma_{\textrm{SS}}\propto\tau_{\perp}\propto1/n$
\cite{Ferreira14,Milletari_2}. The only matrix elements contributing
to the vertex renormalization are those proportional to $\gamma_{1}$
and $\gamma_{13}$. We thus decompose the vertex part in Fig.~(\ref{fig:vertexc}.b)
as $\delta v_{x}=\delta v_{x}^{1}\,\gamma_{1}+\delta v_{x}^{2}\,\gamma_{13}$.
Solving the respective Bethe-Salpeter equation, and taking the trace
of $\delta v_{x}$ together with $\gamma_{1}$ or $\gamma_{13}$,
we obtain $\tilde{v}_{x}=(v+\delta v_{10}+n\,\delta v_{11})\,\gamma_{1}+(\delta v_{20}+n\,\delta v_{22})\,\gamma_{13}$.
For details on the functions $\delta v_{ij}$ refer to SM~\cite{SM}.
Substituting the bare vertex in Eq.~\eqref{eq:bubble} with the renormalized
one, the SH conductivity, in the noncrossing approximation, and to
leading order in $n$ reads 
\begin{align}
\sigma_{\textrm{SH}} & =\frac{\epsilon\,\delta v_{20}}{2\,n\,v\,\eta}+\Big\{\frac{\epsilon\,\delta v_{22}+2\,(v+\delta v_{10})\,\bar{\eta}}{2\,v\,\eta}\nonumber \\
 & -\delta v_{20}\left(\frac{1}{\pi v}+\frac{\bar{\eta}\,m}{2\,v\,\eta^{2}}\right)\Big\}\equiv\mathcal{S}(\epsilon)/n+\mathcal{Q}_{\textrm{nc}}(\epsilon)\,,\label{eq:FullSH}
\end{align}
the main result of the paper. The semiclassical $\mathcal{O}(n^{-1})$
contribution is due to SS, whereas the term in brackets, $\mathcal{Q}_{\textrm{nc}}(\epsilon)$,
here referred to as the \emph{anomalous} SH conductivity, has contributions
stemming from several mechanisms as described below. In Fig.~\eqref{fig:sigmaSH},
we plot the SS contribution as a function of the electrostatic potential
for typical dilute impurity density and SOC magnitude. There is a
parametrically wide region where the SH conductivity attains large
Fermi-energy sensitive values. Generally, the SH angle $\gamma=\sigma_{\textrm{sH}}/\sigma_{xx}$
induced by skew scattering has the following scaling $\gamma\propto n/n^{*}$,
where $n^{*}$ is the areal density of (non-SOC) contaminants and
we assumed $n\ll n^{*}$ (in the opposite limit, $\gamma$ is independent
of $n$). This shows that the SH angle increases linearly with the
SOC impurity density in disordered samples where other mechanisms
limit the charge mobility. The SS contribution is large away from
neutrality, and tends to zero as the impurity scalar energy scale
$\alpha_{0}$ is increased, in agreement with the unitary limit result
of Eq.~\eqref{eq:bubble}. The giant SS contribution to the SH conductivity
has been demonstrated earlier by means of Boltzmann transport theory
\cite{Ferreira14}. However, to our knowledge, a self-consistent treatment
of the spin Hall conductivity, incorporating SS and anomalous processes
on equal footing, had not been reported until now.

\emph{Crossover to the anomalous phase}.\textemdash The anomalous
contribution to the SH conductivity is shown in Fig.~\eqref{fig:sigmaSH}
(dashed lines). It reaches large values of the order of the quantum
of conductance and, contrary to what found for the skew scattering,
it increases as the Fermi energy is lowered. Owing to the $n^{-1}$
scaling of the SS contribution, one would naively expect anomalous
effects to be negligible in the entire dilute regime. Remarkably,
however, a careful inspection of the energy dependence of the spin
Hall conductivity discloses parameter regions where anomalous effects
are dominant in fairly dilute samples, $|\mathcal{Q}_{\textrm{nc}}(\epsilon)|>|\mathcal{S}(\epsilon)/n|$\textemdash see
inset to Fig.~\eqref{fig:sigmaSH}. The rich transport mechanisms
at play in the anomalous ``phase'' are borne out by the distinct
contributions appearing inside brackets in Eq.~\eqref{eq:FullSH}.
In particular, the vertex part associated to the SS ($\delta v_{20}$)
also enters the expression for the anomalous term (traditionally associated
with pure QSJ events). Interestingly, our non-perturbative calculation
shows that diffusion corrections from \emph{reducible} SS diagrams
{[}e.g., diagrams with several ``$Y$s'' in Fig.~\eqref{fig:T-ladder}{]}
strongly renormalize the anomalous term. Consequently,\emph{ even
at the level of a single impurity scattering event}, SS and QSJ cannot
be treated as separate contributions and a correct evaluation of the
anomalous term requires to go beyond the conventional ladder approximation
(see Ref.~\cite{Milletari_2} for details). 

\begin{figure}
\centering{}\includegraphics[width=0.9\columnwidth]{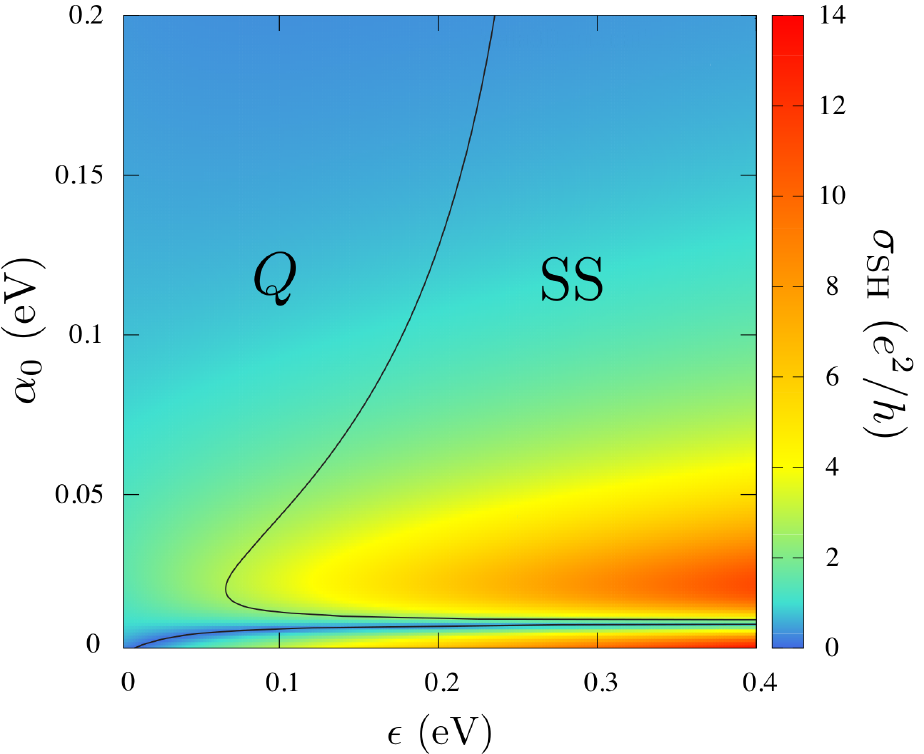} \caption{Phase diagram of the SH conductivity in our model. The diagram shows
the parameter regions in which either $\sigma_{Q}$ or $\sigma_{\textrm{SS}}$
is dominant. The black line is the phase boundary and the different
colors represent the absolute value of $\sigma_{\textrm{SH}}$. We
have used $\alpha_{3}=0.01$ \ eV, $R=4$\ nm and $n=4\times10^{12}$\ cm$^{-2}$. }
\label{fig:transA} 
\end{figure}

The characteristic scalings of the semiclassical SS and anomalous
contributions together with their sharp variation with Fermi energy
provides a smoking gun for an experimental demonstration. In Fig.~\eqref{fig:transA}
we present a representative $\epsilon$ vs $\alpha_{0}$ ``phase
diagram'' of the extrinsic SHE in the intermediate dilute regime,
$n\approx10^{12}$\ cm$^{-2}$, of much experimental relevance. The
black line shows the ``phase boundary'' between a $\mathcal{Q}_{\textrm{nc}}(\epsilon)$-
or $\mathcal{S}(\epsilon)/n$-dominated SHE. The narrow region at
the bottom of the phase diagram corresponds to the special case $|\alpha_{0}|=|\alpha_{3}|$,
for which $\mathcal{S}(\epsilon)/n=0$ irrespectively of $\epsilon$,
c.f. Fig.~\eqref{fig:sigmaSH}. For this particular value, $\mathcal{Q}_{\textrm{nc}}(\epsilon)$
is the only non zero contribution, hence the particular shape of the
phase boundary. Since our calculations are based on a rather conservative
model for the impurity resonance, and thermal effects do not destroy
the robustness of the extrinsic SHE in graphene~\cite{Ferreira14},
the anomalous contributions described here are likely to contribute
to non local signals of recent SH experiments \cite{SOC_G_Balakrishnan_13,SOC_G_Balakrishnan_14,SOC_G_Avsar_14,SOC_G_Wang_Morpurgo_15,SOC_G_Mendes_15}. 

\emph{Summary}. In this work we unveiled an anomalous quantum regime
of the extrinsic spin Hall effect in disordered graphene. Our microscopic
theory\textemdash based on a powerful non-perturbative treatment of
the Kubo\textendash Streda formula\textemdash predicts an experimentally
accessible crossover from skew scattering- to quantum processes-dominated
spin transport, a finding of fundamental importance to the spin Hall
and related effects not envisaged until now. Our work opens the exciting
new prospect of probing quantum spin transport phenomena through (non-local)
electrical measurements in graphene and related heterostructures. 

\emph{Acknowledgements}. M.M. thanks R. Raimondi and G. Vignale for
stimulating discussions. M.M. acknowledges support from the Singapore
National Research Foundation under its fellowship program (NRF Award
No. NRF-NRFF2012-01). A.F. gratefully acknowledges the financial support
of the Royal Society (U.K.) through a Royal Society University Research
Fellowship. 

\begin{center}
\textbf{\large{}\newpage{}{{Supplemental Material}}}
\par\end{center}{\large \par}

\setcounter{equation}{0}

In this supplemental material, we present additional details on the
evaluation of the vertex corrections at the $T$~Matrix level and
give explicit expressions for the functions appearing in the main
text. \\

Response functions determine the transport properties of an electronic
system. In general, the former are expressed as products of two or
more Green's functions of the excited system. In the context of linear
response theory, one usually deals with the product of a Retarded
(R) and an Advanced (A) Green function. A perturbation such as disorder,
not only modifies the individual Green functions but also the response
function itself. The self energy corrections of the individual Green
functions encode the impurity-mediated mean field potential perceived
by the quasiparticles. This information characterises the system at
equilibrium. In order to understand how disorder affects the response
of the system, one needs to look at fluctuations around the mean field
solution. In the diagrammatic language, these fluctuations are encoded
in the 4-point function, or vertex part. Disorder enters the interaction
vertex in the form of repeated incoherent and coherent impurity scattering
processes. Here we focus on the incoherent processes, giving rise
to diffusive corrections to the charge and spin transport. \\

Consider the renormalized interaction vertex, Fig.(1.b) of the main
text. This can be decomposed as $\tilde{v}_{x}=v_{x}+\delta v_{x}$;
here $v_{x}$ is the bare interaction vertex (i.e. in the absence
of disorder) and $\delta v_{x}$ encodes the multiple, incoherent
scattering processes. It is generally convenient to separate the effect
of a single impurity and then consider the repeated processes in a
self consistent way. As explained in the main text, in the $T$~matrix
formalism, the single impurity diagram (the ladder) results from an
infinite resummation of scattering events at all order in the impurity
potential strength, see Fig.(2) of the main text. We write $\delta v_{x}=\bar{v}_{x}+n\,R^{4}\sum_{\mathbf{k}}T\,\mathcal{G}_{\mathbf{k}}^{R}\,\delta v_{x}\,\mathcal{G}_{\mathbf{k}}^{A}\,T^{*}$,
where 
\begin{equation}
\bar{v}_{x}=n\,\int\frac{d^{2}k}{(2\pi)^{2}}\left\{ \hat{T}\,\mathcal{G}_{\mathbf{k}}^{R}\,v_{x}\,\mathcal{G}_{\mathbf{k}}^{A}\,\hat{T}^{\star}\right\} =v\,(a\,\gamma_{1}+b\,\gamma_{13})\label{s:Tladder}
\end{equation}
and we have used $T=(\delta\epsilon-\,\imath\,\eta)\,\gamma_{0}+(m-\imath\,\bar{\eta})\gamma_{3}$.
We have defined the two parameters 
\begin{align}
a & \simeq\epsilon\,\frac{\eta_{+}\,\eta_{-}+\epsilon_{+}\,\epsilon_{-}}{4v^{2}(\eta_{+}+\eta_{-})}+n\,f_{a}(\eta_{+},\eta_{-},\epsilon_{+},\epsilon_{-}),\label{s:Tladder1}\\
b & \simeq\epsilon\,\frac{\eta_{+}\,\epsilon_{-}-\eta_{-}\,\epsilon_{+}}{4v^{2}(\eta_{+}+\eta_{-})}+n\,f_{b}(\eta_{+},\eta_{-},\epsilon_{+},\epsilon_{-}),\nonumber 
\end{align}
\begin{widetext} where %
\begin{equation}
f_{a}(\eta_{+},\eta_{-},\epsilon_{+},\epsilon_{-})=\frac{(\eta_{+}+\eta_{-})(\epsilon_{+}\epsilon_{-}+\eta_{+}\eta_{-})-\pi(\eta_{+}-\eta_{-})(\epsilon_{+}\eta_{-}-\eta_{+}\epsilon_{-})}{4\pi v^{2}(\eta_{+}+\eta_{-})},\label{s:fa}
\end{equation}
and 
\begin{equation}
f_{b}(\eta_{+},\eta_{-},\epsilon_{+},\epsilon_{-})=\frac{(\eta_{+}+\eta_{-})(\epsilon_{+}\eta_{-}-\eta_{+}\epsilon_{-})+\pi(\eta_{+}-\eta_{-})(\epsilon_{+}\epsilon_{-}+\eta_{+}\eta_{-})}{4\pi v^{2}(\eta_{+}+\eta_{-})}.\label{s:fb}
\end{equation}
\end{widetext} From equation~\eqref{s:Tladder1} one can see that
the only matrix elements contributing to the vertex renormalization
are those proportional to $\gamma_{1}$ and $\gamma_{13}$. This suggests
the ansatz: $\delta v_{x}=\delta v_{x}^{1}\,\gamma_{1}+\delta v_{x}^{2}\,\gamma_{13}$.
We obtain 

\begin{widetext} 
\begin{align}
\delta v_{x} & =\delta v_{x}^{1}\,\gamma_{1}+\delta v_{x}^{2}\,\gamma_{13}=v\,(a\,\gamma_{1}+b\,\gamma_{13})+n\,\int\frac{d^{2}k}{(2\pi)^{2}}\left\{ \hat{T}\,\mathcal{G}_{\mathbf{k}}^{R}\,(\delta v_{x}^{1}\,\gamma_{1}+\delta v_{x}^{2}\,\gamma_{13})\,\mathcal{G}_{\mathbf{k}}^{A}\,\hat{T}^{\star}\right\} \label{s:Trent}
\end{align}
\end{widetext} Since no new matrix elements are generated at this
stage, the self consistent equation is close. Taking the trace on
both sides of Eq.~\eqref{s:Trent}, together with $\gamma_{1}$ or
$\gamma_{13}$ we obtain 
\begin{equation}
\left(\begin{array}{c}
\delta v_{x}^{1}\\
\delta v_{x}^{2}
\end{array}\right)=\left\{ \mathbb{I}-\left(\begin{array}{cc}
a & -b\\
b & a
\end{array}\right)\right\} ^{-1}\left(\begin{array}{c}
v\,a\\
v\,b
\end{array}\right),\label{eq:system}
\end{equation}
where $\mathbb{I}$ is the identity matrix. In this way one finds
$\tilde{v}_{x}=(v+\delta v_{x}^{1})\,\gamma_{1}+\delta v_{x}^{2}\,\gamma_{13}$.
It is convenient to separate the renormalized vertices into an impurity
density independent and dependent part as: $\delta v_{x}^{1}=\delta v_{10}+n\,\delta v_{11}$
and $\delta v_{x}^{2}=\delta v_{20}+n\,\delta v_{22}$. These are
the vertex parts that appear in the final expression for the spin
Hall conductivity, Eq.~(7) of the main text. Their explicit expressions
are shown below: \begin{widetext} 
\begin{equation}
\delta v_{10}=v\frac{4v^{2}\epsilon(\eta_{+}+\eta_{-})(\epsilon_{+}\epsilon_{-}+\eta_{+}\eta_{-})-\epsilon^{2}\left(\eta_{+}^{2}+\epsilon_{+}^{2}\right)\left(\eta_{-}^{2}+\epsilon_{-}^{2}\right)}{\epsilon^{2}\left(\eta_{+}^{2}+\epsilon_{+}^{2}\right)\left(\eta_{-}^{2}+\epsilon_{-}^{2}\right)-8\epsilon\,v^{2}(\eta_{+}+\eta_{-})(\epsilon_{+}\epsilon_{-}+\eta_{+}\eta_{-})+16v^{4}(\eta_{+}+\eta_{-})^{2}}\,,\label{s:v10}
\end{equation}
\begin{align}
\delta v_{11} & =\frac{v}{\pi}4\,v^{2}(\eta_{+}+\eta_{-})\Big\{16v^{4}(\eta_{+}-\eta_{-})^{2}[\pi(\eta_{+}+\eta_{-})(\eta_{+}\epsilon_{-}-\eta_{-}\epsilon_{+})+(\eta_{+}+\eta_{-})(\eta_{+}\eta_{-}+\epsilon_{+}\epsilon_{-})]-8\epsilon\pi v^{2}(\eta_{+}+\eta_{-})^{2}\nonumber \\
 & \times\left(\eta_{+}^{2}+\epsilon_{+}^{2}\right)\left(\eta_{-}^{2}+\epsilon_{-}^{2}\right)+\epsilon^{2}\left(\eta_{+}^{2}+\epsilon_{+}^{2}\right)\left(\eta_{-}^{2}+\epsilon_{-}^{2}\right)[(\eta_{+}+\eta_{-})(\eta_{+}\eta_{-}+\epsilon_{+}\epsilon_{-})-\pi(\eta_{+}+\eta_{-})(\eta_{+}\epsilon_{-}-\eta_{-}\epsilon_{+})]\Big\}\nonumber \\
 & /\left[\epsilon^{2}\left(\eta_{+}^{2}+\epsilon_{+}^{2}\right)\left(\eta_{-}^{2}+\epsilon_{-}^{2}\right)-8\epsilon\,v^{2}(\eta_{+}+\eta_{-})(\epsilon_{+}\epsilon_{-}+\eta_{+}\eta_{-})+16v^{4}(\eta_{+}+\eta_{-})^{2}\right]^{2}\,,
\end{align}
\begin{equation}
\delta v_{20}=v\frac{4\epsilon v^{2}(\eta_{+}+\eta_{-})(\epsilon_{-}\eta_{+}-\epsilon_{+}\eta_{-})}{\epsilon^{2}\left(\eta_{+}^{2}+\epsilon_{+}^{2}\right)\left(\eta_{-}^{2}+\epsilon_{-}^{2}\right)-8\epsilon\,v^{2}(\eta_{+}+\eta_{-})(\epsilon_{+}\epsilon_{-}+\eta_{+}\eta_{-})+16v^{4}(\eta_{+}+\eta_{-})^{2}}\,,\label{s:v20}
\end{equation}
\begin{align}
\delta v_{22} & =\frac{v}{\pi}4\,v^{2}(\eta_{+}+\eta_{-})\Big\{16v^{4}(\eta_{+}-\eta_{-})^{2}[\pi(\eta_{+}-\eta_{-})(\eta_{+}\eta_{-}+\epsilon_{+}\epsilon_{-})+(\eta_{+}^{2}+\eta_{-}^{2})(\eta_{+}\epsilon_{-}-\eta_{-}\epsilon_{+})]-8\epsilon\pi v^{2}(\eta_{+}-\eta_{-})\nonumber \\
 & \times\left(\eta_{+}^{2}+\epsilon_{+}^{2}\right)\left(\eta_{-}^{2}+\epsilon_{-}^{2}\right)+\epsilon^{2}\left(\eta_{+}^{2}+\epsilon_{+}^{2}\right)\left(\eta_{-}^{2}+\epsilon_{-}^{2}\right)[(\eta_{+}+\eta_{-})(\eta_{+}\epsilon_{-}-\eta_{-}\epsilon_{+})+\pi(\eta_{+}-\eta_{-})(\eta_{+}\eta_{-}+\epsilon_{+}\epsilon_{-})]\Big\}\nonumber \\
 & /\left[\epsilon^{2}\left(\eta_{+}^{2}+\epsilon_{+}^{2}\right)\left(\eta_{-}^{2}+\epsilon_{-}^{2}\right)-8\epsilon\,v^{2}(\eta_{+}+\eta_{-})(\epsilon_{+}\epsilon_{-}+\eta_{+}\eta_{-})+16v^{4}(\eta_{+}+\eta_{-})^{2}\right]^{2}\,.
\end{align}
\end{widetext}
\end{document}